\documentclass{aa}  
\usepackage[varg]{txfonts}
\usepackage[normalem]{ulem}  
\usepackage{color}
\usepackage{url}
\usepackage{hyperref}

\usepackage{upgreek}
\usepackage{graphicx}   
\usepackage{amsmath}    
\usepackage{amssymb}    
\usepackage{xcolor}
\usepackage{comment}

\bibpunct{(}{)}{;}{a}{}{,} 
                   
\newcommand{\rs}{R_{\rm S}}
\newcommand{\msun}{M_{\odot}}
\newcommand{\rmd}{{d}}
\newcommand{\upi}{\pi}

\begin{document}

\title{Accurate analytic formula for light bending in Schwarzschild metric}
 \titlerunning{Approximate light bending formula}
   
\author{Juri Poutanen\inst{1,2,3}
}
\authorrunning{Juri Poutanen}

\institute{
Department of Physics and Astronomy,  FI-20014 University of Turku, Finland \\ \email{juri.poutanen@utu.fi}
\and Space Research Institute of the Russian Academy of Sciences, Profsoyuznaya str. 84/32, 117997 Moscow, Russia 
\and Nordita, KTH Royal Institute of Technology and Stockholm University, Roslagstullsbacken 23, SE-10691 Stockholm, Sweden
}


\abstract
{We propose new analytic formulae describing light bending in Schwarzschild metric. 
For emission radii above the photon orbit at 1.5 Schwarzschild radius, the formulae have an accuracy of better than 0.2\% for the bending angle and 3\% for the lensing factor for any trajectories that turn around a compact object by less  than about 160\degr. 
In principle, they can be applied to any emission point above the horizon of the black hole. 
The proposed approximation can be useful for problems involving emission from neutron stars and accretion discs around compact objects when fast accurate calculations of light bending are required. 
It can also be used to test the codes that compute light bending using exact expressions via elliptical integrals. }


\maketitle



\section{Introduction}
 
Understanding physical processes in the vicinity of black holes (BHs) and neutron stars (NSs) requires detailed treatment of light propagation from a compact source to the distant observer. 
In a general case of a rotating compact object, this is a complex, numerically extensive problem \citep[e.g.][]{Dexter16,NP:18}.
For a slowly rotating object, the Schwarzschild metric can be used, but even in this case numerical, time-consuming evaluations of elliptical integrals to describe light bending is needed. 
The situation becomes acute when one needs to fit the data with a model varying many parameters which may require thousands, if not millions, of iterations.  
Such a problem exists, for example, when trying to determine NS parameters from the pulse form observed from millisecond pulsars which have oblate shape \citep{Miller15,Watts16RvMP,Bogdanov19L26,Riley19,MillerLamb19}.

\begin{figure}
\centering
\includegraphics[width=.8\columnwidth]{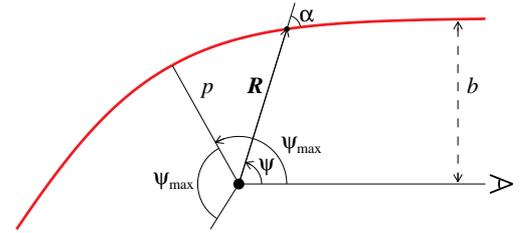} 
\caption{Geometry of light bending in Schwarzschild metric.  The observer is situated on the right at $\psi=0$. 
\label{fig:bend_geom}}
\end{figure}

In many applications the position of the emission point is defined, e.g. by the radius-vector $\vec{R}$ of the emission point and the azimuthal angle $\psi$ vector $\vec{R}$ makes with the direction to the observer  (see Fig.~\ref{fig:bend_geom}).  
We then need to compute the emission angle $\alpha$ that the photon trajectory makes with $\vec{R}$. 
For that we would have to tabulate  $\psi(\alpha)$ at a grid of radii $R$, then reverse the dependence to $\alpha(\psi)$ and finally interpolate in the resulting tables to find $\alpha$ for given $R$ and $\psi$.  
An analytical formula for $\alpha(R,\psi)$ would simplify and speed up calculations.
It can also be used for testing other more accurate routines for light bending. 

A powerful approximation to the bending integral in Schwarzschild metric of the required form $\alpha(R,\psi)$ was discovered by \citet{Beloborodov:02}. 
He showed that there is a nearly linear relation between $x= 1-\cos\alpha$ and $y=1- \cos\psi$: 
\begin{equation}\label{eq:ABappr} 
x = 1-\cos\alpha \approx (1-u)y = (1-u)(1- \cos\psi) ,  
\end{equation}
where $u=\rs /R$ is the compactness, $\rs=2GM/c^2$ is the Schwarzschild radius of the central object of mass  $M$.   
This approximation has high accuracy for direct trajectories (i.e. not passing through the turning point, i.e. periastron) and not very compact star, with radius exceeding $2\rs$. 
A useful property of this approximation, however, is that it is linear in three parameters: $\cos\alpha$, $\cos\psi$, and $u$. 
Thus, for any known two parameters, the third can be found easily. 
For example, if we are interested in the total bending angle corresponding to a given compactness, we fix $\cos\alpha=0$, find $\psi_{\max}$ from a simple relation $\cos\psi_{\max}=-u/(1-u)$ and the total bending angle as $2\psi_{\max}-\pi$.  
This approximation can be also used to obtain an approximate form of the photon trajectory for the given impact parameter (which depends on $\alpha$ and $u$, see Eq.\,(\ref{eq:impact}) below), as given by Eq.\,(3) in \citet{Beloborodov:02}. 
Using similar approach other approximate forms for the photon trajectory and the total bending angle were suggested by \citet{Semerak2015}. 
 
In this paper, however, we are interested only in a simple approximation for $\alpha(R,\psi)$. 
We propose the following approximation: 
\begin{equation} \label{eq:bending_new}
x = (1-u) \,y \left\{1 + \frac{u^2y^2}{112}- \frac{e}{100} u y \left[\ln\left(1-\frac{y}{2}\right) +\frac{y}{2}\right]  \right\},
\end{equation}
where $e$ is the base of the natural logarithm.
It works for trajectories that make less than half of full turn around central object and for the radii all the way to the horizon. 
We then compare our new approximation to other  approximations proposed in the literature and test it on two well-known problems: the light curve from two antipodal hotspots at a NS and the line emission from the accretion disc around a Schwarzschild BH.

\section{Light bending in Schwarzschild metric} 
  \label{sec:bending}

 \subsection{Bending angle}

Consider photon passing near a gravitating centre (BH or  NS) and escaping to infinity (see Fig.~\ref{fig:bend_geom}). 
In Schwarzschild metric the shape of  photon's trajectory  is described by the equation \citep[][p. 673]{MTW73}
\begin{equation}\label{eq:motion}
\left( \frac{1}{R^2} \frac{\rmd R}{\rmd \psi} \right)^2 + \frac{1}{R^{2}}
(1-u) = \frac{1}{b^2}, 
\end{equation}
where $R$ is the circumferential radius, $\psi$ is the azimuthal angle, $b$ is the impact parameter.  
The impact parameter and the angle, $\alpha$, between the radial direction and the photon trajectory are related by \citep[e.g.][]{Beloborodov:02}
\begin{equation}\label{eq:impact}
  b=\frac{R}{\sqrt{1-u}} \sin\alpha .
\end{equation}

\begin{figure}
\centering
\includegraphics[width=0.85\columnwidth]{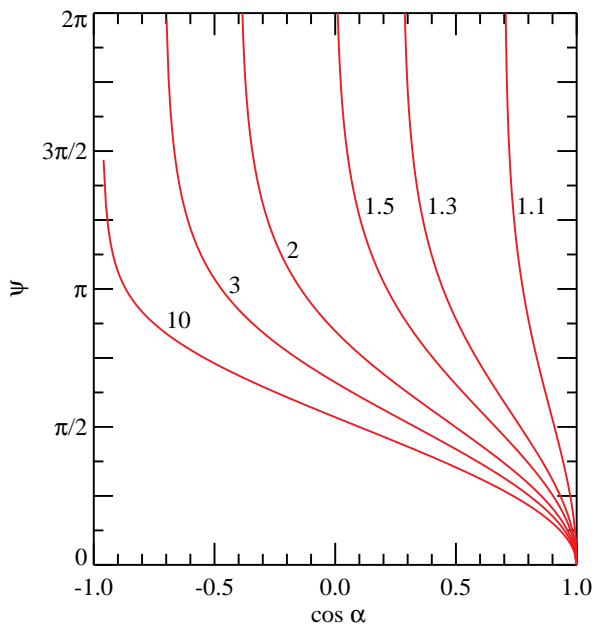} 
 \caption{Light bending relation between the cosine of the emission angle $\alpha$ and the angle  $\psi$ between the line of sight and the radius-vector of the emission point   computed using exact relations (\ref{eq:bend1})--(\ref{eq:bend2}) for Schwarzschild metric for six different  emission radii $R/\rs=1.1, 1.3, 1.5, 2, 3$ and 10, marked next to corresponding curves.   
\label{fig:psi_of_alpha}
 }
\end{figure}

\begin{figure}
\centering
\includegraphics[width=0.85\columnwidth]{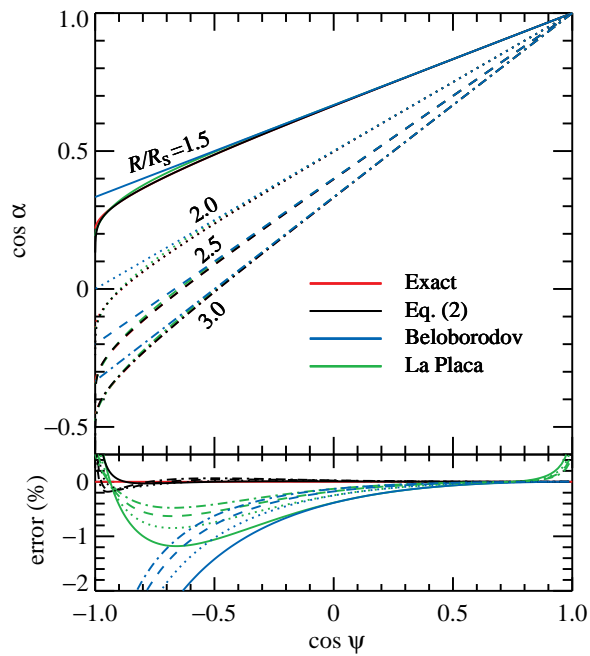} 
\caption{\textit{Upper panel}: Light bending relation between the cosine of the emission angle $\alpha$ and the cosine of the angle  $\psi$ between the line of sight and the radius-vector of the emission point  in Schwarzschild metric. 
The red curves give the exact relation.
Our  new approximate relation (\ref{eq:bending_new}) is shown with the black curves. 
The blue straight lines are for the \citet{Beloborodov:02} approximation (\ref{eq:ABappr}), while the green curves represent approximation (\ref{eq:laplaca}) by \citet{LaPlaca19}. 
The red, green and black curves practically coincide. 
The solid, dotted, dashed, and dot-dashed  curves correspond to radii $R/\rs=1.5, 2, 2.5, 3$, respectively. 
\textit{Bottom panel}: the relative error in the emission angle $\delta\alpha/\alpha$ for three approximations as compared to the exact result. 
Same notations as in the upper panel. 
\label{fig:bending}
}
\end{figure}

In a BH case, a photon with impact parameter $b\leq b_{\rm cr}=\rs \ 3\sqrt{3}/2$  \citep[][p. 675]{MTW73} may be captured by the central object. 
The critical impact parameter $b_{\rm cr}$ corresponds to the critical emission angle
\begin{equation}\label{eq:capture}
\alpha_{\rm cr}=\arcsin (3\sqrt{3}\ u\ \sqrt{1-u}\  / 2) . 
\end{equation} 
If emission radius is small $R\leq 1.5\rs$ (i.e. $u\geq 2/3$), only photons with  $\alpha\leq\alpha_{\rm cr}$ can escape to infinity.  
For larger emission radius $R>1.5\rs$, all photons with $\alpha\leq\pi/2$ escape.
In these cases, the observer angle $\psi(R,\alpha)$, i.e. the angle between the radius vector of the emission point and the photon momentum at infinity, is given by the integral \citep[e.g.][]{PFC83,Beloborodov:02}
\begin{equation}\label{eq:bend1}
  \psi  (R,\alpha)=\int_R^{\infty} \frac{\rmd r}{r^2} \left[ \frac{1}{b^2} -
       \frac{1}{r^2}\left( 1- \frac{\rs}{r}\right)\right]^{-1/2},
\end{equation}
with $b$ given by Eq.\,(\ref{eq:impact}). 

If $R>1.5\rs$, the critical emission angle is instead $\pi- \alpha_{\rm cr}$, and the condition for photon capture can be written as 
\begin{equation}\label{eq:capture2}
\alpha> \pi- \alpha_{\rm cr} > \pi/2,
\end{equation}
or 
\begin{equation}\label{eq:capture_cos}
\cos \alpha< - \sqrt{ 1 - \frac{27}{4} u^2\ (1-u) } . 
\end{equation}
Thus, photons emitted at angle $\pi/2<\alpha\leq\pi-\alpha_{\rm cr}$ escape, but first they pass though the turning point (see Fig.~\ref{fig:bend_geom}) at azimuthal angle 
\begin{equation}\label{eq:psimax}
\psi_{\max}=\psi (p,\pi/2).
 \end{equation}  
The periastron, $p$, can be found by setting $\rmd R/\rmd\psi = 0$ in Eq.\,(\ref{eq:motion}) and solving the resulting cubic equation $p^3=b^2 (p-\rs)$ to get
\begin{equation}
p=-\frac{2}{\sqrt{3}} \ b \cos\left\{[\arccos(b_{\rm cr}/b)+2\pi]/3\right\}.
\end{equation}
The observer angle is then given by 
\begin{equation} \label{eq:bend2}
\psi(R,\alpha)= 2\psi_{\max} - \psi(R,\pi-\alpha).
\end{equation}
A numerical method to accurately compute bending integrals is described, for example, by \citet{SNP18}. 
The resulting relation between $\psi$ and $\cos\alpha$ for different radii is shown in Fig.~\ref{fig:psi_of_alpha}. 
We see that $\psi$ diverges when $\cos\alpha$ approaches critical values. 
This corresponds to many rotations of a photon around the BH and may result in multiply images. 

For majority of realistic astrophysical situations, we can limit ourselves only to the primary image with $\psi<\pi$, because other images may be blocked by the accretion disc and the flux decreases rapidly with the number of turns \citep{Luminet79}. 
In case of a NS, the trajectories that pass through the stellar surface will be truncated. 
For a spherical star, this means that we will be interested only in trajectories with $\cos\alpha>0$. 
If a NS is rapidly rotating, its shape is not spherical anymore and, in principle, some trajectories with $\cos\alpha<0$ may also become possible. 
For the primary image, the dependence $\cos\alpha(\cos\psi)$ would be sufficient and  we plot it in  Fig.\,\ref{fig:bending}.

\begin{figure}
\centering
\includegraphics[width=0.9\columnwidth]{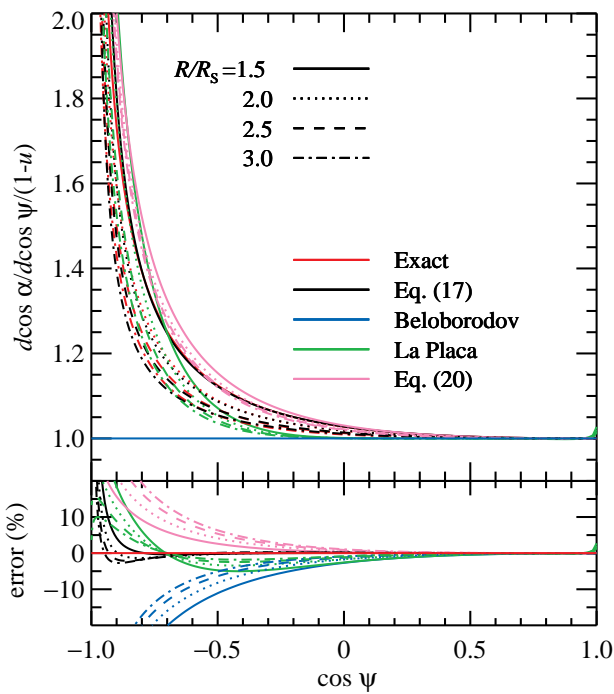} 
 \caption{Same as Fig.\,\ref{fig:bending}, but for the lensing factor ${\cal D}$. Approximation (\ref{eq:lensing_series}) is shown by pink curves. 
   }
\label{fig:lensing}
\end{figure}

\begin{figure*}
\centering
\includegraphics[width=0.8\columnwidth]{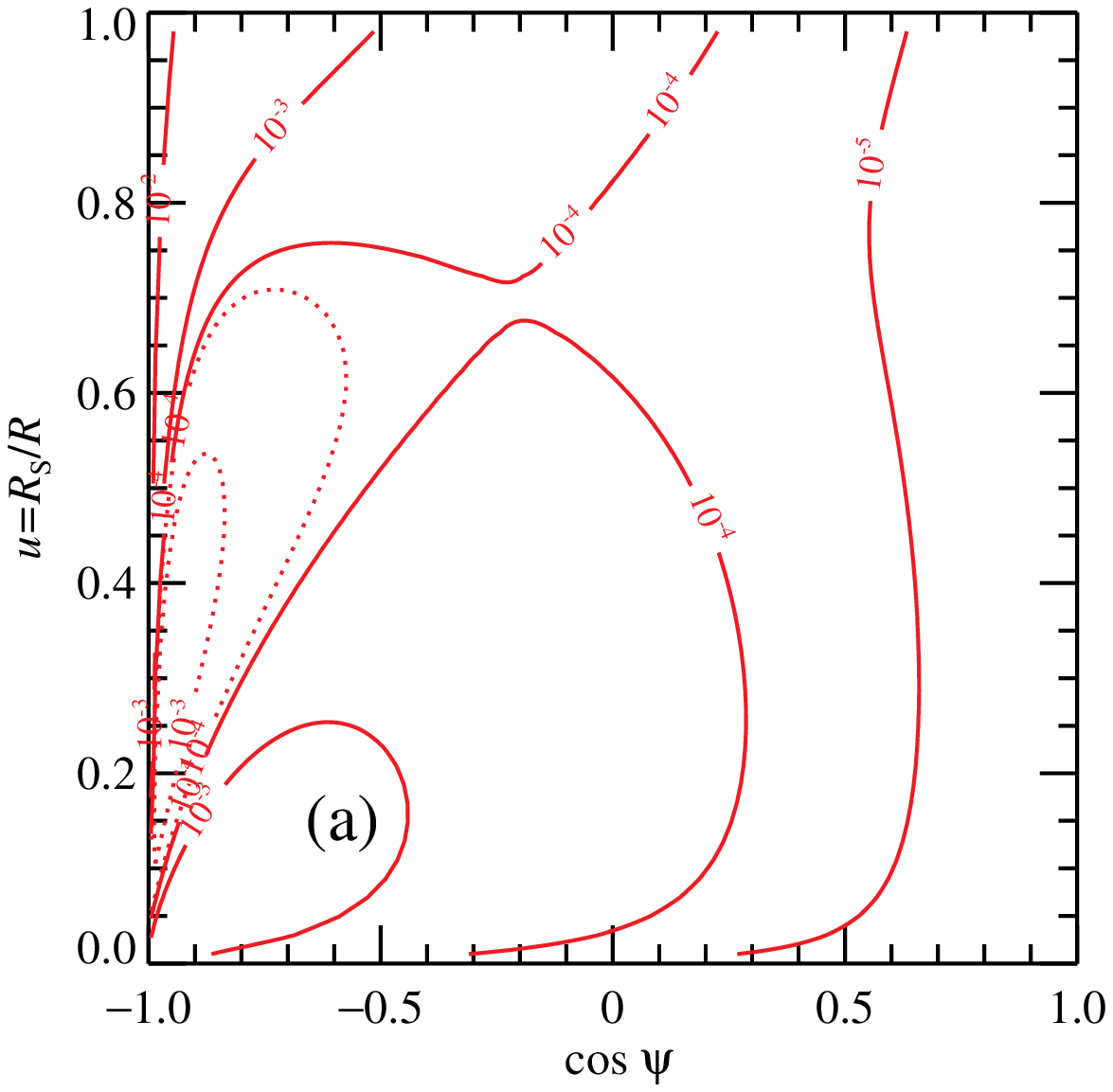} \hspace{1cm}
\includegraphics[width=0.8\columnwidth]{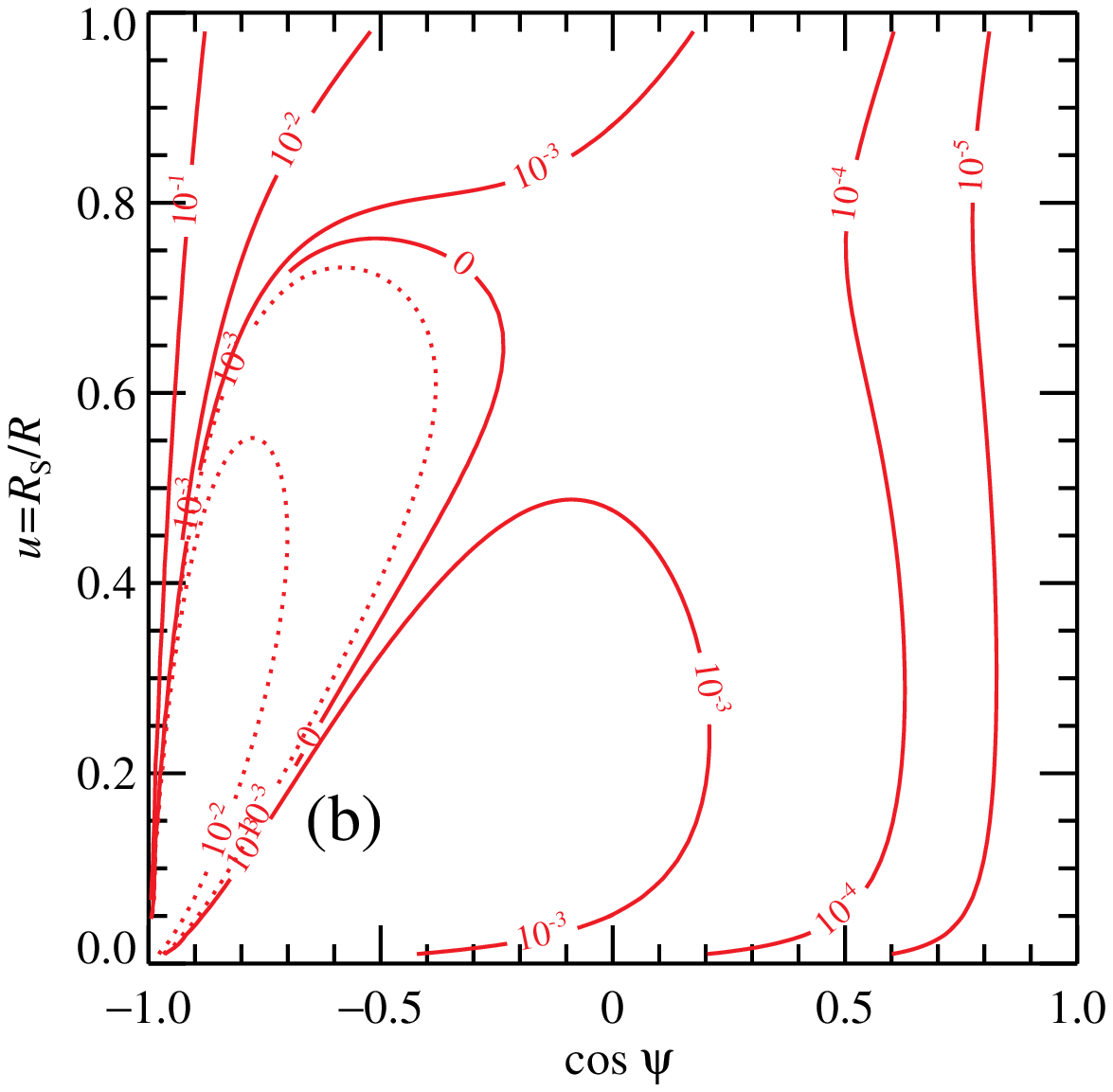} 
\caption{Contours of the constant relative error on (a) the bending angle $\delta\alpha/\alpha$ and 
(b) the lensing factor for our approximations given by Eqs.\,(\ref{eq:bending_new}) and (\ref{eq:deriv_bending_new}). 
Neighbouring contours differ by a factor of 10 in the value of the error.
Solid and dotted curves represent positive and negative deviations, respectively. 
}
\label{fig:errors}
\end{figure*}

\subsection{Lensing factor} 

Now we turn to a problem of evaluating flux from a  surface element of area $dS$. 
Without losing a generality, we can assume that the normal to the surface is along the radial direction $\vec{R}$.  
The flux observed from this element is proportional to the product of the radiation intensity $I$ and the solid angle occupied by the element on the observer's sky $d\Omega$. 
The solid angle can be represented via the impact parameter 
as 
\begin{equation}
d\Omega = \frac{b\, db\, d\phi}{D^2}, 
\end{equation}
with  $D$ being the distance to the source and $\phi$ is the azimuthal angle in the spherical coordinate system with the $z$-axis directed along the line of sight. 
Expressing the element area as  $dS= R^2 d\cos\psi\,d\phi$ and using Eq.\,(\ref{eq:impact}) we get \citep{Beloborodov:02} 
\begin{equation} \label{eq:omega}
d\Omega =   \frac{dS}{D^2} 
 \frac{b }{R^2} \left|  \frac{db}{d\cos\psi} \right| = \frac{dS \cos\alpha}{D^2}    \frac{1}{1-u} \frac{d\cos\alpha}{d\cos\psi}. 
\end{equation}
We see that the solid angle has two terms: the first is just the solid angle that the element observed at inclination $\alpha$ would occupy in flat space $dS\,\cos\alpha/D^2$, while the second factor corrects for light bending. 
Thus in calculations of the observed flux, it is not only important to get an accurate estimate of the emission angle $\alpha$ for a given $\psi$, but also to evaluate accurately the lensing factor  
\begin{equation} \label{eq:lensing}
{\cal D} =   \frac{1}{1-u} \frac{d\cos\alpha}{d\cos\psi} ,
\end{equation} 
which is shown in Fig.\,\ref{fig:lensing}.

\section{Approximate light bending formulae}

We need to design approximations of the form $\alpha(u,\psi)$ and  ${\cal D}(u,\psi)$. 
A simple approximate relation (\ref{eq:ABappr})  discovered by  \citet{Beloborodov:02} is not very accurate for  large emission angles $\alpha$ and large compactness $u\gtrsim1/2$. 
This is demonstrated in Fig.\,\ref{fig:bending}, where  \citet{Beloborodov:02} approximation (blue lines) is compared with the exact relation (red curves). 
We see that the error on the emission angle $\delta\alpha/\alpha$ grows systematically with decreasing $\cos\psi$ (i.e. increasing $\psi$, which corresponds to the emission points further from our line of sight).  
For small compactness, e.g. $u\lesssim1/3$ (i.e. $R\gtrsim 3 \rs$), and the NS case, it is not a problem, because we  are mostly interested in trajectories with $\cos\alpha>0$, where the error does not exceed 0.7\%. 
The error grows, however, with compactness and for $u=1/2$ it is already 10\%.

The situation is even worse for the lensing factor (\ref{eq:lensing}). 
Equation (\ref{eq:ABappr}) implies ${\cal D}=1$, while the exact value grows rapidly at negative $\cos\psi$ (see Fig.\,\ref{fig:lensing}), e.g. at 
 $\cos\psi=-0.7$ (i.e. $\psi=134\degr$), deviation from unity exceeds 10\% for $u=1/3$ and 15\% for $u=1/2$. 
It is thus clear that the approximation may introduce significant error in the flux observed, for example, from a spot at the far side of a NS or from the accretion disc viewed at large inclination. 
Realization of this problem motivates us to look for a different, more accurate approximation. 
 
Approximation (\ref{eq:ABappr}) was derived by \citet{Beloborodov:02} from the exact expression of the bending angle (\ref{eq:bend1}) by expanding the integral in Taylor series over small parameter $x$ and obtaining a new Taylor series for $y(x)$. 
\citet{PB06} got an expression for the reverse relation: 
\begin{equation} \label{eq:PB06_app}
x = (1-u) y \left( 1 + \frac{u^2}{112} y ^2 \right)  , 
\end{equation}
which, however, still has the same problems as the original approximation (\ref{eq:ABappr}), because deviations appear at large values of the argument $y$. 

Recently, a purely phenomenological formula was proposed by \citet{LaPlaca19}: 
\begin{equation} \label{eq:laplaca}
x = (1-u) y \left\{  1 + k_1 u [ 1-\cos(\psi-k_2)]^{k_3} \right\}  , 
\end{equation}
where $k_1=0.1416$, $k_2=1.196$ and $k_3=2.726$. 
This approximation is shown in Fig.\,\ref{fig:bending} by the green curves. 
We see that it is better than 1\% accurate for most of the angles of interest. 
However, it does not reproduce well the exact behaviour at small angles $\psi \approx\alpha/\sqrt{1-u}$ having there unphysical jumps, which are also reflected in the jumps in the derivative (lensing factor) at small $\psi$ (see green curves in Fig.\,\ref{fig:lensing}). 
The lensing factor has a typical accuracy of 3--5\% and deviates by more than 5\% from the exact values at $\cos\psi\lesssim -0.8$. 

We instead suggest to design a fitting formula that keeps the correct asymptotic behaviour at $\psi\rightarrow 0$ as  given by Eq.~(\ref{eq:PB06_app}), but at the same time provides a sufficient curvature when $\cos\psi$ is close to $-1$ (i.e. $y=2$). 
For that we add a logarithmic term of the type $\propto \ln(1-y/2)$ that satisfies the second condition, but subtract the terms of the corresponding Taylor expansion around $y=0$ in order to satisfy the first condition. 
We found that a good fit to the exact bending relation is provided by Eq.~(\ref{eq:bending_new}). 
It gives an error below 0.06\%  for $\cos\psi>-0.5$  (i.e. for the angle $\psi<120\degr$ from the radial direction) and any radius exceeding $1.5\rs$. 
At these radii, the error exceeds 0.2\%  only for $\cos\psi<-0.95$, i.e. $\psi>162\degr$ (see black curves in Fig.\,\ref{fig:bending}) corresponding to the emission points behind the compact object. 
The contours of constant errors on the plane $(u,\cos\psi)$ are shown in  Fig.\,\ref{fig:errors}a.
We see that approximation works rather well even for radii between event horizon and the photon orbit, $\rs<R<1.5\rs$ (i.e. $2/3<u<1$), of course, only for emission angles very close to the radial direction, so that the photon trajectory makes less than half of the full turn around a compact object. 
 
The lensing factor  implied by Eq.\,(\ref{eq:bending_new}),
\begin{equation}\label{eq:deriv_bending_new}
{\cal D} =  1 + \frac{3u^2y^2}{112}  - \frac{e}{100} u y \left[ 2\, \ln\left(1-\frac{y}{2}\right) + y\frac{1-3y/4}{1-y/2}\right] ,
\end{equation}
also has high accuracy. 
Fig.\,\ref{fig:errors}b shows the contours of constant error on the plane $(u,\cos\psi)$. 
We see that the error exceeds 10\% only for $\cos\psi<-0.9$ and $u>0.8$. 
For an object with radius exceeding the photon orbit, i.e.  $u<2/3$, the error is below 0.3\% for $\cos\psi>-0.5$ 
(see also black curves in Fig.\,\ref{fig:lensing}). 

Another way to approximate the lensing factor (\ref{eq:lensing}) is to start from its following form 
\begin{equation}\label{eq:lensing_deriv}
{\cal D} =  \frac{1}{1-u}  \frac{\sin\alpha}{\sin\psi} 
\frac{1}{\cos\alpha}  \frac{d\sin\alpha}{d\psi} .
\end{equation}
The derivative $d\psi/d\sin\alpha$ can be written in an implicit form following from Eq.\,(\ref{eq:bend1}) as  
\begin{equation}\label{eq:deriv_psi}
\frac{d\psi} {d\sin\alpha} = 
\frac{R}{\sqrt{1-u}} 
\int_R^{\infty} \frac{\rmd r}{r^2} \left[ 1 -
       \frac{b^2}{r^2}\left( 1- \frac{\rs}{r}\right)\right]^{-3/2} .
\end{equation} 
Expanding it as well as $\sin\alpha$ and $\cos\alpha$ in Eq.\,(\ref{eq:lensing_deriv}) in small parameter $x=1-\cos\alpha$ up to $x^2$ but keeping factor $\sin\psi$ in the denominator, we get\footnote{A similar approach for the lensing factor was used by \citet{DeFalco2016}. That paper, however, has a number of flaws: in calculations of the bending angle for $\alpha>\pi/2$, $\psi_{\max}$ was computed as $\psi(R,\alpha=\pi/2)$ instead of the correct  $\psi_{\max}=\psi(p,\alpha=\pi/2)$, see Eq.\,(\ref{eq:bend2}); the expression for the solid angle (proportional to our lensing factor) contains an excessive factor $\sin\alpha/\sin\psi$; there is an error in Eq.\,(30), where $1-C...$ should be $-1/2-C...$ instead; and the pulse profiles from a hotspot on a rapidly rotating NS in their Fig.\,9 have unphysical jumps before eclipses instead of going to zero.}
\begin{equation}\label{eq:lensing_series}
{\cal D} \approx \frac{\sqrt{2y}}{\sin\psi}    
\left[ 1- \frac{y}{4} + y^2 \left( -\frac{1}{32} + \frac{5}{224} u^2 \right)  \right] .
\end{equation}
Here in the final expression we used Eq.\,(\ref{eq:ABappr}) and substituted $y=x/(1-u)$ to get ${\cal D}$ as a function of $\psi$, not $\alpha$. 
The factor $\sin\psi$ in the denominator gives rise to a diverging behavior at $\psi\rightarrow\pi$ (see Fig.\,\ref{fig:lensing}) and allows to describe the actual dependence of the lensing factor slightly better than just a constant ${\cal D}=1$ from Belobodorov's approximation, but much worse than other approximations considered above. 
 
\section{Applications}

\subsection{Hotspots at a neutron star surface}

\begin{figure*}
\centering
\includegraphics[width=0.8\columnwidth]{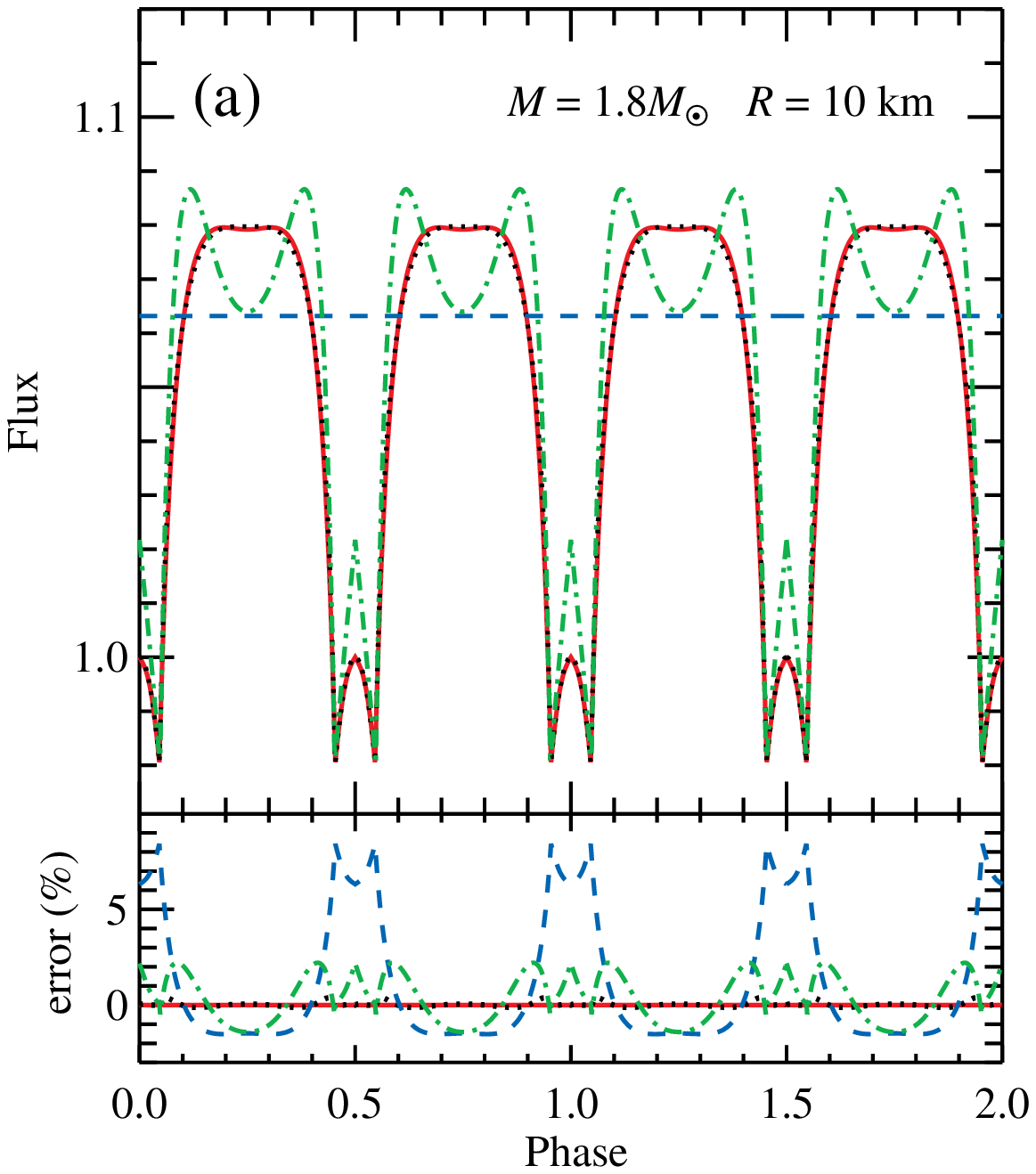} \hspace{1cm}
\includegraphics[width=0.8\columnwidth]{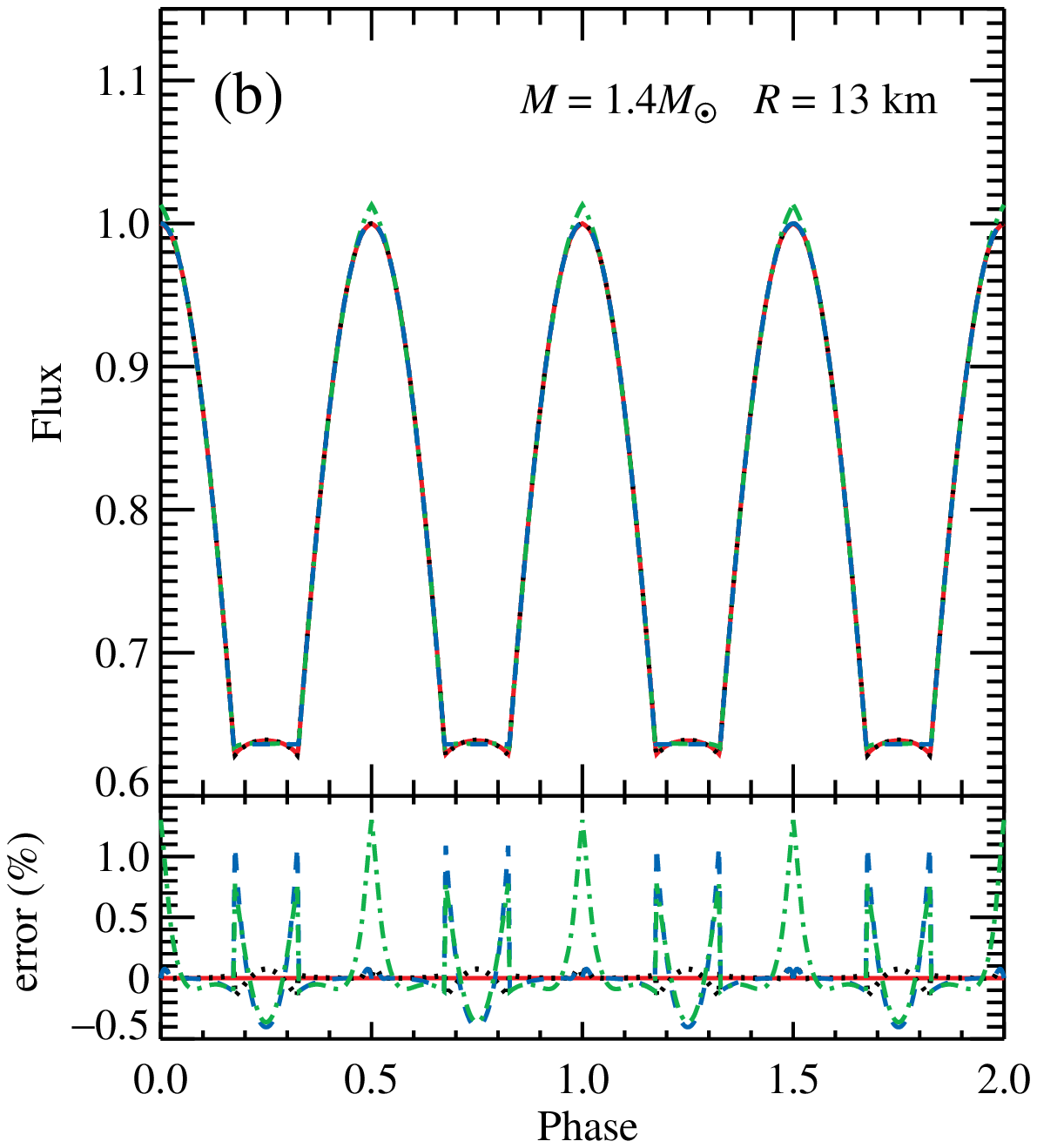} 
\caption{Scaled flux as a function of pulsar phase produced by two antipodal hotspots at the surface of a NS  for two different compactnesses: 
(a) $M=1.8\msun$, $R$=10 km; (b)  $M=1.4\msun$, $R$=13 km.   
Both the observer inclination and the magnetic obliquity are fixed at 90\degr. 
The red solid curves give the results of exact calculations  of bending. 
Our approximation  (given by Eqs.\,(\ref{eq:bending_new}) and (\ref{eq:deriv_bending_new}) is shown with black dotted curves. 
The blue dashed and green dot-dashed curves correspond to the approximations by \citet{Beloborodov:02} and \citet{LaPlaca19}, respectively. 
The lower subpanels show the relative error in the flux for the same three approximations of light bending compared to the exact result.    }
\label{fig:pulsar}
\end{figure*}

Let us now consider a test case which demonstrates the power of approximate formulae for light bending. 
We consider two antipodal spots of area $dS$ at a slowly rotating spherical NS of radius $R$ and mass $M$. 
Let the observer unit vector be $\hat{\vec{o}}=(\sin i,0,\cos i)$ and the co-latitude of the primary be $\theta$. 
The unit-vector corresponding to the radius vector of the primary hotspot varies with rotational phase $\varphi$ as $\hat{\vec{R}}=(\sin\theta\cos\varphi,\sin\theta\sin\varphi, \cos\theta)$. 
This gives us  the expression for the angle between $\hat{\vec{o}}$ and $\hat{\vec{R}}$:  
\begin{equation} 
\cos\psi=\hat{\vec{o}} \cdot \hat{\vec{R}} = \cos i \cos\theta + \sin i \sin\theta\cos\varphi. 
\end{equation}
For the secondary spot, we substitute $\varphi\rightarrow \varphi+\pi$ and $\theta \rightarrow \pi-\theta$.
The observed bolometric flux is $F=I\,d\Omega$,
where the solid angle given by Eq.\,(\ref{eq:omega}). 
Thus the flux is \citep{Beloborodov:02}
\begin{equation} 
F= I \frac{dS}{D^2}\ {\cal D}\  \cos\alpha  . 
\end{equation}
If the intensity  at the NS surface is angle-independent, the pulse profile is fully determined by variation of ${\cal D}\cos\alpha$. 
Thus we plot in Fig.\,\ref{fig:pulsar} the sum of the scaled fluxes ${\cal D}\cos\alpha$ from two spots situated at the equator for the equatorial observer ($\theta=i=90\degr$). 
This geometry maximizes the range of angles $\psi$. 
Our approximation gives accuracy of 0.37\% for a  compact NS   ($M=1.8\msun$ and $R=10$\,km giving $u=0.53$), while for a smaller compactness ($M=1.4\msun$ and $R=13$\,km,  $u=0.32$) the accuracy is 0.15\%. 
The  \citet{LaPlaca19} approximation is  2.2\%  and  1.3\% accurate and the  \citet{Beloborodov:02} approximation gives an error of 8.4\%  and 1.1\% for the two considered cases. 
 
\subsection{Line profile from an accretion disc}

\begin{figure}
\centering
\includegraphics[trim={0 4cm 0 3cm},clip,width=0.95\columnwidth]{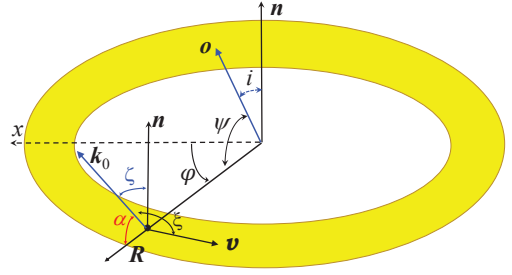} 
\caption{Geometry of emission from an accretion disc ring. }
\label{fig:geometry_disc}
\end{figure}

Let us now consider a problem of line emission from a Keplerian accretion disc around a Schwarzschild BH as discussed, for example, by \citet{Chen89} and \citet{Fabian89}.
We compute the line profile seen by observers at different inclinations $i$ along direction $\hat{\vec{o}}=(\sin i,0,\cos i)$.
We define a coordinate system with the $z$-axis normal  to the disc $\hat{\vec{n}}=(0,0,1)$, so that  the disc lies in the equatorial plane $\theta=\upi/2$. 
The radius-vector of an element of the disc surface at azimuthal angle $\varphi$, $\hat{\vec{R}}=(\cos\varphi, \sin\varphi,0)$ makes  angle $\psi$ to the line of sight (see Fig.~\ref{fig:geometry_disc} for geometry):
\begin{equation}\label{eq:cospsi}
 \cos\psi=\hat{\vec{R}} \cdot \hat{\vec{o}} =  \sin i\ \cos\varphi.
\end{equation}

Because in Schwarzschild metric the photon trajectories are planar, the direction of the photon momentum close to the disc surface can be described by a unit vector 
\begin{equation}\label{eq:k0}
\hat{\vec{k}}_0=[ \sin\alpha\ \hat{\vec{o}} +\sin(\psi-\alpha)\ \hat{\vec{R}}]/\sin\psi, 
\end{equation}
where $\cos\alpha=\hat{\vec{k}}_0 \cdot \hat{\vec{R}}$.
The surface element at (circumpherential) radius $R$ is moving with Keplerian velocity $\vec{v}=v(-\sin\varphi,\cos\varphi,0)$  with  $\beta=v/c = \sqrt{u/2(1-u)}$ relative to a static observer at this radius \citep[see e.g.][]{Luminet79}.
The corresponding Lorentz factor  is 
\begin{equation}\label{eq:lorentz}
\gamma=\frac{1}{\sqrt{1-\beta^2}}= \sqrt{ \frac{1-u}{1-3u/2}} . 
\end{equation}
The photon momentum makes angle $\xi$ with the velocity vector 
\begin{equation}\label{eq:cosxi2}
\cos\xi=\hat{\vec{v}} \cdot \hat{\vec{k}}_0
=\frac{\sin\alpha}{\sin\psi} \hat{\vec{v}} \cdot \hat{\vec{o}}=
- \frac{\sin\alpha}{\sin\psi}\sin i\ \sin\varphi\   ,
\end{equation}
and with the local disc normal it makes angle $\zeta$: 
\begin{equation}\label{eq:coszeta}
  \cos\zeta =\hat{\vec{n}} \cdot \hat{\vec{k}}_0 = \frac{\sin\alpha}{\sin\psi} \hat{\vec{n}} \cdot \hat{\vec{o}} =  \frac{\sin\alpha}{\sin\psi} \cos i  . 
\end{equation}
The Doppler factor  is 
\begin{equation}\label{eq:dop}
\delta=\frac{1}{\gamma(1-  {\vec{\beta}}  \cdot \hat{\vec{k}}_0)} = \frac{1}{\gamma(1-\beta\cos\xi)} .
\end{equation}
From Lorentz transformation one can get the angle that photon momentum makes with the local normal in the comoving frame \citep[see e.g.][]{PG03,PB06}
\begin{equation}\label{eq:cosalphadop}
\cos\zeta'= \delta\cos\zeta .
\end{equation}

The specific flux observed from a surface element at photon energy $E$ is
\begin{equation} \label{eq:dF_E}
 \rmd F_E=I_E\ \rmd\Omega,
\end{equation}
where $I_E$ is the specific intensity of radiation at infinity, which is related to that in the comoving disc element frame  
\begin{equation}
  I_{E} = \left (\frac{E}{E'}\right )^3 I'_{E '} (\zeta') 
\end{equation}
and the energy ratio \citep{Luminet79,Chen89}
\begin{equation} \label{eq:EEpr}
\frac{E}{E'}=\delta \sqrt{1-u} =  \frac{\sqrt{1-3u/2} } {1+\beta\sin i\sin\phi\sin\alpha/\sin\psi } 
\end{equation}
combines the effects of the gravitational redshift and Doppler effect. 
The solid angle occupied by the surface element of area $dS=R dR d\varphi/\sqrt{1-u}$ is given by equation similar to (\ref{eq:omega}): 
\begin{equation}\label{eq:omega_disc}
\rmd \Omega=\frac{\rmd S \cos \zeta}{D^2} \frac{1}{1-u} \frac{\rmd\cos\alpha}{\rmd\cos\psi}.
\end{equation}
The observed spectral flux (Eq.~\ref{eq:dF_E}) now reads
\begin{equation}\label{eq:fluxspot}
d F_{E} (R,\varphi) =(1-u)^{3/2} \delta^{3} I'_{E'}(\zeta')
 \frac{\rmd S  \cos\zeta }{D^2} {\cal D}. 
\end{equation}
The observed flux from the disc is then obtained by integrating Eq.~(\ref{eq:fluxspot}) over radius and azimuthal angle
\begin{equation}\label{eq:flux_azimuth_ave}
F_E =  \frac{1}{D^2}  
 \int  (1-u)\,R \rmd R\ \int_0^{2\pi} \!  \!  \!  \!  \! \rmd \varphi  \, 
 \delta^{3} I'_{E'}(\zeta') {\cal D} \cos\zeta . 
\end{equation}
Inside the integrand, for a given $R$ and $\varphi$ (and given inclination $i$) we compute $\psi$ using Eqs.~(\ref{eq:cospsi}). 
It is used to get $\alpha$  and ${\cal D}$ using approach described in Sect.\,\ref{sec:bending}. 
Then $\xi$ and $\zeta$  can be computed from Eqs.~(\ref{eq:cosxi2}) and  (\ref{eq:coszeta}), respectively. 
Using the Keplerian velocity and the Lorentz factor given by Eq.~(\ref{eq:lorentz}), we get then the Doppler factor $\delta$ from Eq.\,(\ref{eq:dop}). 
Furthermore, from Eqs.~(\ref{eq:cosalphadop}) and (\ref{eq:EEpr}), we get the photon zenith angle in the comoving frame $\zeta'$ and the comoving energy $E'$, which are needed for obtaining  $I'_{E'}(\zeta')$.

\begin{figure}
\centering
\includegraphics[width=0.85\columnwidth]{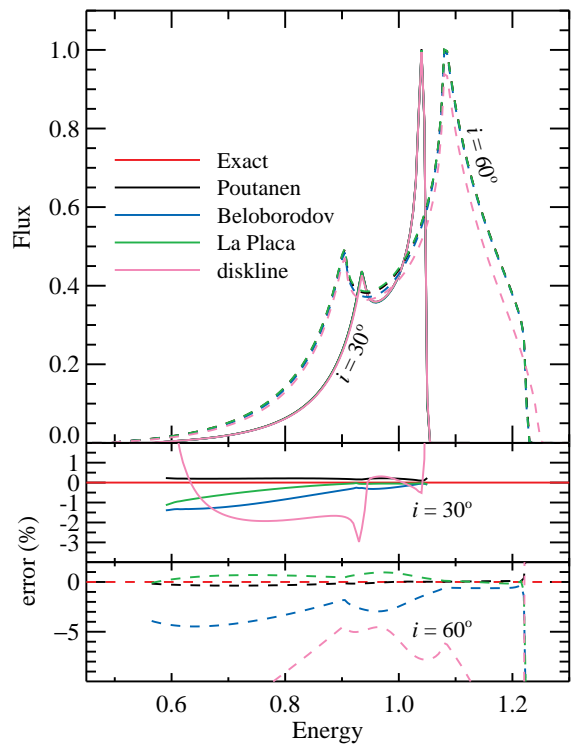} 
\caption{\textit{Upper panel}:  Profiles of the emission line from an accretion disc ring extending from 3 to 50$\rs$ around a Schwarzschild BH (or a slowly rotating NS) with the emissivity radial dependence $\propto R^{-2}$. 
The solid and dashed curves are for the observer inclination 30$\degr$ and 60$\degr$, respectively. 
The red curves corresponds to the exact treatment of bending. 
The results using our new approximation  given by Eqs.\,(\ref{eq:bending_new}) and (\ref{eq:deriv_bending_new}) is shown with the black curves. 
The blue and green curves correspond to the approximations of \citet{Beloborodov:02} and \citet{LaPlaca19},   respectively. 
All the curves overlap. 
The pink curves show the profile with no bending accounted for (as in the {\sc xspec} model {\sc diskline}). 
The profiles are renormalized by a factor giving maximum of unity for the exact profile. 
\textit{Bottom panel}: the relative error in the line flux for the considered approximations.    }
\label{fig:line}
\end{figure}

As an example, we consider a case with isotropic emission in a narrow line centered at comoving energy $E_0=1$ with width $\sigma=2\times10^{-3}$ 
from an accretion disc ring extending from 3 to 50$\rs$ with the radial dependence of the emissivity $\propto R^{-2}$ as was assumed in the original publication by \citet{Fabian89}. 
The line profiles observed at two inclinations using exact treatment of light bending and different approximations are shown in Fig.\,\ref{fig:line}.
We see that our approximation gives accuracy better than 0.4\%, while other proposed approximations give errors from 1 to 5\%. 
Ignoring the light bending, as was done in the well-known {\sc xspec} \citep{Arn96} model {\sc diskline}  \citep{Fabian89}, gives an error  that grows from 2\% at $i=30\degr$ to
20\% at $i=60\degr$. 

We note that our approximation is nearly independent of the emission radius. 
For example, if the line is produced in a narrow ring at 3$\rs$, our approximation gives accuracy of 0.13\%  and 1.5\% for  $i=30\degr$  and  $60\degr$, respectively, while 
the corresponding errors are 2.7\% and 14\% for the \citet{Beloborodov:02} approximations and 1\% and 2.8\% for the \citet{LaPlaca19} approximation. 
The  {\sc diskline}  model, on the other hand, has a typical error of 5\% and 15\%, respectively, but it rises  sharply towards the line peaks reaching there 40\% and 70\%. 

\citet{2011MNRAS.414.1269W}  showed that the line profiles  from accretion discs mostly depend in the inner disc radius (which is the function of the black hole spin),  while the effect of the spin on photon trajectories is minor. 
Because our approximation works equally well for emission radii well within 3$\rs$ (see Fig.\,\ref{fig:errors}), it can,  in principle, be used for calculations of the line profiles from the discs around rotating black holes too.  
Detailed calculations are left for future work.   

\section {Summary}
\label{sec:summary} 
 
In this paper we proposed new approximation for light bending in Schwarzschild metric. 
It can be applied to any emission point above the horizon of the BH and also for trajectories that pass through the turning point, but make less than half of full turn. 
For emission radii above the photon orbit at 1.5 Schwarzschild radius, the approximation has an accuracy of better than 0.2\% for the bending angle and 3\% for the lensing factor for photon orbits turning by less than 160\degr\ around a compact object.
This approximation can be useful for problems involving rotating oblate NSs and accretion disc around compact object when fast accurate calculations of light bending are required. 
The proposed formulae can be also used to check the results of exact calculations.

\section*{Acknowledgments} 
This research has been supported by the grant 14.W03.31.0021 of the Ministry of Science and Higher Education of the Russian Federation.  
I thank Joonas N\"attil\"a and Dmitry Yakovlev for comments. 
 

\label{lastpage}
\end{document}